\documentclass[a4paper,12pt]{article}
\usepackage{amsmath,amssymb} \usepackage{euler,eucal}
\usepackage[utf8]{inputenc}
\usepackage[T1]{fontenc}
\usepackage[english]{babel}
\usepackage{tgpagella}
\usepackage[unicode=true,plainpages=false]{hyperref}
\usepackage{graphicx}
\hypersetup{colorlinks=false,pdfborder={0 0 0.1},linkcolor=magenta,anchorcolor=magenta,urlcolor=blue,citecolor=blue,
          pdftitle={Exact NMR simulation of protein-size spin systems using tensor train formalism},
          pdfauthor={D. V. Savostyanov, S. V. Dolgov, J. M. Werner, Ilya Kuprov},
          pdfkeywords={density matrix renormalization group, alternating minimal energy, tensor train, nuclear magnetic resonance, protein}
          }

\usepackage[left=30mm,right=30mm,top=20mm,bottom=30mm,a4paper]{geometry}
\usepackage{algorithmic}
\usepackage[ruled]{algorithm}
\usepackage{amsthm}
\theoremstyle{definition}

\def\O{\mathcal{O}}
\def\eps{\varepsilon}
\def\im{\mathrm{i}}
\def\trace{\mathop{\mathrm{Tr}}\nolimits}
\def\tru{\star}
\def\Zeeman{{\mathrm{Z}}}

\usepackage[version=3]{mhchem}
\def\cH{\ce{^{1}H}}
\def\cC{\ce{^{13}C}}
\def\cN{\ce{^{15}N}}

\def\hat{\mathaccent"705E }
\def\hrho{\hat\rho}
\def\hx{\hat x}
 \def\hhA{\hat{\hat A}} \def\mA{A} \def\An{\mathbf{A}^{(n)}}
\def\hH{\hat H} \def\hhH{\hat{\hat H}}
\def\hJ{\hat J}
\def\hO{\hat O}
\def\hS{\hat S} \def\hSn{\hat S^{(n)}}
\def\hvS{\hat{\vec{S}}} \def\hvSn{\hat{\vec{S}}^{(n)}}
\def\hs{\hat\sigma} %\def\hsn{\hat{\sigma}^{(n)}}
\def\hhR{\hat{\hat R}}
\def\vB{\vec{B}_0}
\def\Id{\mathbf{1}}
\begin{document}
\author{
  D. V. Savostyanov\footnotemark[2],~
  S. V. Dolgov\footnotemark[4],~
  J. M. Werner\footnotemark[3],~
  Ilya Kuprov\footnotemark[2]
}
\title{Exact NMR simulation of protein-size spin systems using tensor train formalism}
\date{June 18, 2014}
\maketitle
\renewcommand{\thefootnote}{\fnsymbol{footnote}}
\footnotetext[2]{University of Southampton, School of Chemistry, Highfield Campus, Southampton SO17 1BJ, United Kingdom  ({\tt d.savostyanov@soton.ac.uk})}
\footnotetext[3]{University of Southampton, Centre for Biological Systems, Highfield Campus, Southampton SO17 1BJ, United Kingdom}
\footnotetext[4]{Max-Planck Institute for Mathematics in the Sciences, Inselstrasse 22, Leipzig 04103, Germany}
\renewcommand{\thefootnote}{\arabic{footnote}}
\maketitle

\begin{abstract}
We introduce a new method, based on alternating optimization, for compact representation of spin Hamiltonians and solution of linear systems of algebraic equations in the tensor train format. We demonstrate the method's utility by simulating, without approximations, a \cN~NMR spectrum of ubiquitin --- a protein containing several hundred interacting nuclear spins. Existing simulation algorithms for the spin system and the NMR experiment in question either require significant approximations or scale exponentially with the spin system size. We compare the proposed method to the \emph{Spinach} package that uses heuristic restricted state space techniques to achieve polynomial complexity scaling. When the spin system topology is close to a linear chain (e.g. for the backbone of a protein), the tensor train representation is more compact and can be computed faster than the sparse representation using restricted state spaces.
\end{abstract}

\section{Introduction} \label{sec:intro}  %%%%%%%%%%%% INTRO
The amount of patience required to simulate exactly a nuclear magnetic resonance (NMR) spectrum of an $N$-spin system scales approximately as $\O(2^N)$. That much is rarely available, and considerable thought has consequently been given over the last decade to more efficient methods~\cite{nmr-simpson-2002,emsley-2009,dumont-1997,spinach-2011,spinevolution-2006}, particularly those that promise to achieve that objective in polynomial time. Such algorithms do exist~\cite{emsley-2009,karabanov-acc-2011}, but they make significant \emph{a priori} assumptions about the spin system evolution --- it is usually assumed that the system stays weakly correlated for the duration of the experiment~\cite{karabanov-acc-2011,kuprov-2007}.

Outside the NMR community, significant progress was recently made with the development of tensor structured methods~\cite{schollwock-2005,kolda-review-2009,schollwock-2011,khor-survey-2011,hackbusch-2012}, all of which descend broadly from the density matrix renormalization group (DMRG)~\cite{white-dmrg-1992,white-dmrg-1993} as well as matrix product state (MPS)~\cite{fannes-mps-1992,klumper-mps-1993} and matrix product operator (MPO)~\cite{verstraete-mpo-2004} formalisms. Typical applications of DMRG in condensed matter theory are 1D spin chains~\cite{schollwock-2005,bursill-dmrgT-1996,shibata-dmrgT-1997,wang-dmrgT-1997} with recent extensions to 2D lattices~\cite{liang-dmrg2d-1994,mcculloch-dmrg2d-2001,xiang-dmrg2d-2001,white-dmrg2d-2012}. DMRG has also been put to good use in electronic~\cite{white-abinitio-1999,chan-dmrg-2002,zgid-spinsym-2008,kurashige-abinitio-2009,wouters-finitefield-2012,sharma-spin-2012,chemps2-2014} and nuclear~\cite{pittel-nuclear-2006,rotureau-open-2006} structure theory, but magnetic resonance spectroscopy has so far received little attention --- the spin systems encountered in the daily practice of NMR and EPR (proteins, radicals, polynucleotides, polysaccharides) are irregular three-dimensional room-temperature networks with multiple interlocking loops in the spin coupling graph and no identical couplings~\cite{cavanagh-nmr-2007}. When the strict requirement for correct wavefunction phase during the very long (milliseconds to seconds) dissipative spin system trajectories is added to the list, time-domain DMRG methods are currently struggling.

There are some biologically relevant cases, however, that may still be treated as linear chains --- for the purposes of simulating simple backbone NMR experiments, protein side chains may often be ignored. This makes the corresponding spin system a weakly branched linear chain that is amenable to DMRG type treatment. Simple NMR experiments can also be reformulated as a matrix-inverse-times-vector problem in the frequency domain, for which efficient algorithms in tensor product formats have recently emerged~\cite{jeckelmann-dmrgsolve-2002,DoOs-dmrg-solve-2011,ds-amr1-2013,ds-amr2-2013}. We report in this communication the behavior of the AMEn algorithm~\cite{ds-amr1-2013,ds-amr2-2013,ds-dmrgamen-2014}, applied to the solution of the NMR simulation problem in the frequency domain, as well as to the technical task of adding together, without loss of accuracy, tensor train representations of thousands of spin Hamiltonian terms for a protein.

Having integrated the algorithms described below into \emph{Spinach} (a large-scale magnetic resonance simulation library~\cite{spinach-2011}), we are reporting here the first exact quantum mechanical simulation of a liquid-state 1D NMR spectrum for a protein backbone spin system with several hundred coupled spins. Beyond the physical assumptions made by chemists at the problem formulation stage and the controllable numerical rounding error of the tensor train format itself~\cite{osel-tt-2011}, there are no approximations.

\section{Tensor product formats} \label{sec:ttdef}  %%%%%%%%%%%%%%%%%%% TT DEF
Tensor product expressions appear naturally in spin dynamics because the state space of a multi-spin system is a direct product of state spaces of individual spins~\cite{ernst-nmr-1987}. A simple example is the nuclear Zeeman interaction Hamiltonian
\begin{equation}\label{eq:HZ}
 \hH_\Zeeman = \sum_{n=1}^N \vB \:\cdot\: \An \:\cdot\: \hvSn,
\end{equation}
where
 $N$ is the number of spins,
 $\vB$ is the applied magnetic field,
 $\An$ are nuclear chemical shielding tensors,
 and the sum runs over all nuclei.
Cartesian components of nuclear spin operators
$
\hvSn = \begin{bmatrix} \hSn_x & \hSn_y & \hSn_z \end{bmatrix}
$
have the following tensor product form
\begin{equation}\label{eq:Sn}
 \hSn_{\{x,y,z\}} = \Id \otimes  \cdots \otimes \Id \otimes \hs_{\{x,y,z\}} \otimes \Id \otimes \cdots  \otimes \Id,
\end{equation}
where
 $\Id$ denotes an identity matrix of appropriate dimension and
 Pauli matrices $\hs_x, \hs_y, \hs_z$ occur at the $n$-th position in the tensor product sequence.
This representation is known in numerical linear algebra as the canonical polyadic (CP) format~\cite{kolda-review-2009}.
Although CP representations have been known in magnetic resonance spectroscopy for a long time~\cite{ernst-cogwheel-1997},
  they suffer in practice from rapid inflation ---
  spin Hamiltonians encountered in NMR and ESR (electron spin resonance) systems can be complicated~\cite{ernst-nmr-1987}
  and, even for simple initial conditions, the number of terms in the canonical decomposition increases rapidly during system evolution.
More ominously, the number of CP terms can change dramatically after small perturbations of the Hamiltonian or the system state.
A simple example is
\begin{equation}\label{eq:S}
 \hS_z
     = \sum_{n=1}^N \hSn_z
     = \lim_{\eps\to 0} \frac{(\Id+\eps\hs_z)^{\otimes N} - \Id^{\otimes N}}{\eps},
\end{equation}
where
$
\hat a^{\otimes N} = \hat a \otimes \cdots \otimes \hat a.
$
The left hand side of this equation contains $N$ direct product terms, given by Eq.~\eqref{eq:Sn},
  but the expression approximating it on the right hand side has only two direct product terms,
  and one could be tempted to use it to reduce storage and CPU time.
However, both terms of the approximation grow to infinity when $\eps\to 0,$
  and the accuracy is lost due to rounding errors.
Such instabilities in the CP format make it difficult to use ---
   in finite precision arithmetic the number of terms in the decomposition quickly becomes equal to the dimension of the full state space and any efficiency savings disappear.

Unlike the CP format, which is an \emph{open} tensor network, \emph{closed} tensor network formats are stable to small perturbations.
The most popular closed tensor network format was repeatedly rediscovered and is currently known under three different names:
  DMRG in condensed-matter physics~\cite{white-dmrg-1992,white-dmrg-1993},
  MPS~\cite{fannes-mps-1992,klumper-mps-1993,verstraete-faith-2006,PerezGarcia-mps-2007}/MPO~\cite{verstraete-mpo-2004,crosswhite-longrange-2008} in computational physics, and
  TT (tensor train) in numerical linear algebra~\cite{osel-tt-2011}.
A tensor train is defined, using the standard notation of numerical linear algebra~\cite{dkos-eigb-2014}, as follows:
\begin{equation}\label{eq:tt}
   \hx  = \tau(\hx^{(1)},\ldots,\hx^{(N)})
  = \sum_{\alpha_1,\ldots,\alpha_{N-1}}
         \hx_{\alpha_1}^{(1)} \otimes
         \hx_{\alpha_1,\alpha_2}^{(2)} \otimes\cdots\otimes
         %\hx_{\alpha_{N-2},\alpha_{N-1}}^{(N-1)} \otimes
         \hx_{\alpha_{N-1}}^{(N)}.
\end{equation}
The TT representation of the total $\hS_z$ operator in Eq.~\eqref{eq:S} is similar to the high--dimensional Laplacian~\cite{khkaz-lap-2012}:
\begin{equation}\label{eq:ttS}
  \hS_z  = \sum_{\alpha_1=1}^2 \ldots \sum_{\alpha_{N-1}=1}^2
      \hH_{\alpha_1}^{(1)} \otimes \hH_{\alpha_1,\alpha_2}^{(2)} \otimes\cdots\otimes \hH_{\alpha_{N-1}}^{(N)},
\end{equation}
with
  $
  \hH^{(1)}  = \begin{bmatrix} \hs_z & \Id \end{bmatrix},
  $
  $
  \hH^{(2)}  = \ldots=\hH^{(N-1)} = \begin{bmatrix} \Id & 0 \\ \hs_z & \Id \end{bmatrix},
  $
  and
  $
  \hH^{(N)} = \begin{bmatrix} \Id & \hs_z  \end{bmatrix}^\top.
  $
The number of terms in each summation (known as \emph{bond dimension}, or \emph{TT rank}) is two, and the number of entries of the decomposition is now bounded.
The TT representation of $\hS_z$ in Eq.~\eqref{eq:ttS} has $4N-4$ single--spin operators, each of which is 
  either zero, or 
  identity $\Id$, or 
  the Pauli matrix $\hs_z.$
The CP representation of $\hS_z$ in Eq.~\eqref{eq:S} has $N^2$ such operators --- the tensor train representation is clearly more memory--efficient.

Another notable example is the ZZ coupling Hamiltonian that often makes an appearance in models of simple linear spin chains:
\begin{equation}\label{eq:J}
 \hJ = \sum_{m>n} \hS^{(n)}_z \hS^{(m)}_z.
\end{equation}
As written, this is a CP format with $N(N-1)$ terms and $N^2(N-1)$ single-spin operators entering direct products.
The corresponding TT representation is
\begin{equation}\label{eq:ttJ}
  \begin{split}
  \hJ &  = \sum_{\alpha_1=1}^3 \ldots \sum_{\alpha_{N-1}=1}^3
      \hJ_{\alpha_1}^{(1)} \otimes \hJ_{\alpha_1,\alpha_2}^{(2)} \otimes\cdots\otimes \hJ_{\alpha_{N-1}}^{(N)},
  \\
  \hJ^{(1)}  = \begin{bmatrix} 0 & \hs_z & \Id \end{bmatrix}, &
  \qquad
  \hJ^{(2)}  = \ldots = \hJ^{(N-1)} = \begin{bmatrix} \Id & 0 & 0 \\ \hs_z & \Id & 0 \\ 0 & \hs_z & \Id \end{bmatrix},
  \qquad
  \hJ^{(N)} = \begin{bmatrix} \Id \\ \hs_z \\ 0 \end{bmatrix}.
 \end{split}
\end{equation}
Here each summation runs over three terms only, and the total number of single-spin operator matrices appearing in $\hJ^{(n)}_{\alpha_{n-1},\alpha_n}$ is $9N-12$, much fewer than the one of the CP format in Eq.~\eqref{eq:J}.

Storage requirements of tensor structured representations (both CP and TT)
    stand in sharp contrast with the classical approach to magnetic resonance simulations~\cite{nmr-simpson-2002,spinevolution-2006},
    where the Hamiltonian is represented as a $2^N\times 2^N$ sparse matrix with all non-zero entries stored in memory.
As soon as the matrix is assembled, CPU and memory resources grow exponentially with the number of spins $N,$ making the simulation prohibitively difficult for large systems.
Tensor structured methods avoid this problem (it is known colloquially as \emph{the curse of dimensionality}) by keeping all data in compressed formats of the form given in Eqs.~\eqref{eq:HZ} and~\eqref{eq:tt} and manipulating it without ever opening up the Kronecker products.

A very considerable body of literature exists on manipulating expressions directly in tensor product formats~\cite{kolda-review-2009,khor-survey-2011,hackbusch-2012}.
In particular, a given matrix may be converted into the TT format using sequential singular value decompositions~\cite{schollwock-2005,PerezGarcia-mps-2007,osel-tt-2011}.
Given tensors in the TT format, 
   one can perform linear or bilinear operations (addition, element-wise multiplication, matrix-vector multiplication)~\cite{schollwock-2011,osel-tt-2011},
  Fourier transform~\cite{browne-fft-2007,dks-ttfft-2012},
  and convolution~\cite{khkaz-conv-2013} directly in the TT format,
  avoiding exponentially large arrays and computational costs.

These developments would have permitted large--scale magnetic resonance simulations entirely in the TT format,
   were it not for a significant obstacle ---
  the summation operation in tensor train representations is an expensive procedure that carries a significant accuracy penalty due to the need to re-compress the representation to keep the bond dimensions low. Spin Hamiltonians of practically interesting biological systems contain many thousands one-- and two--spin terms of the kind shown in Eq.~\eqref{eq:Sn}.
Intermediate expressions in spin dynamics simulations also frequently involve large sums.
We demonstrate below that in those circumstances the standard bundle-and-recompress tensor network summation procedure leads either to the bond dimension expansion beyond the limits of modern computing hardware, or to a catastrophic accuracy loss. This problem also occurs with three--dimensional potentials encountered in electronic structure theory~\cite{khor-ml-2009,sav-rr-2009,gos-kryl-2012}. Here we propose an alternative algorithm for computing large sums, based on alternating tensor train optimization, and use it to enable NMR simulations on protein-size spin systems.

\section{Simulation setting and experimental context} \label{sec:exp} %%%%%%%%%% EXP
\begin{figure}[t]
 \includegraphics[width=\linewidth]{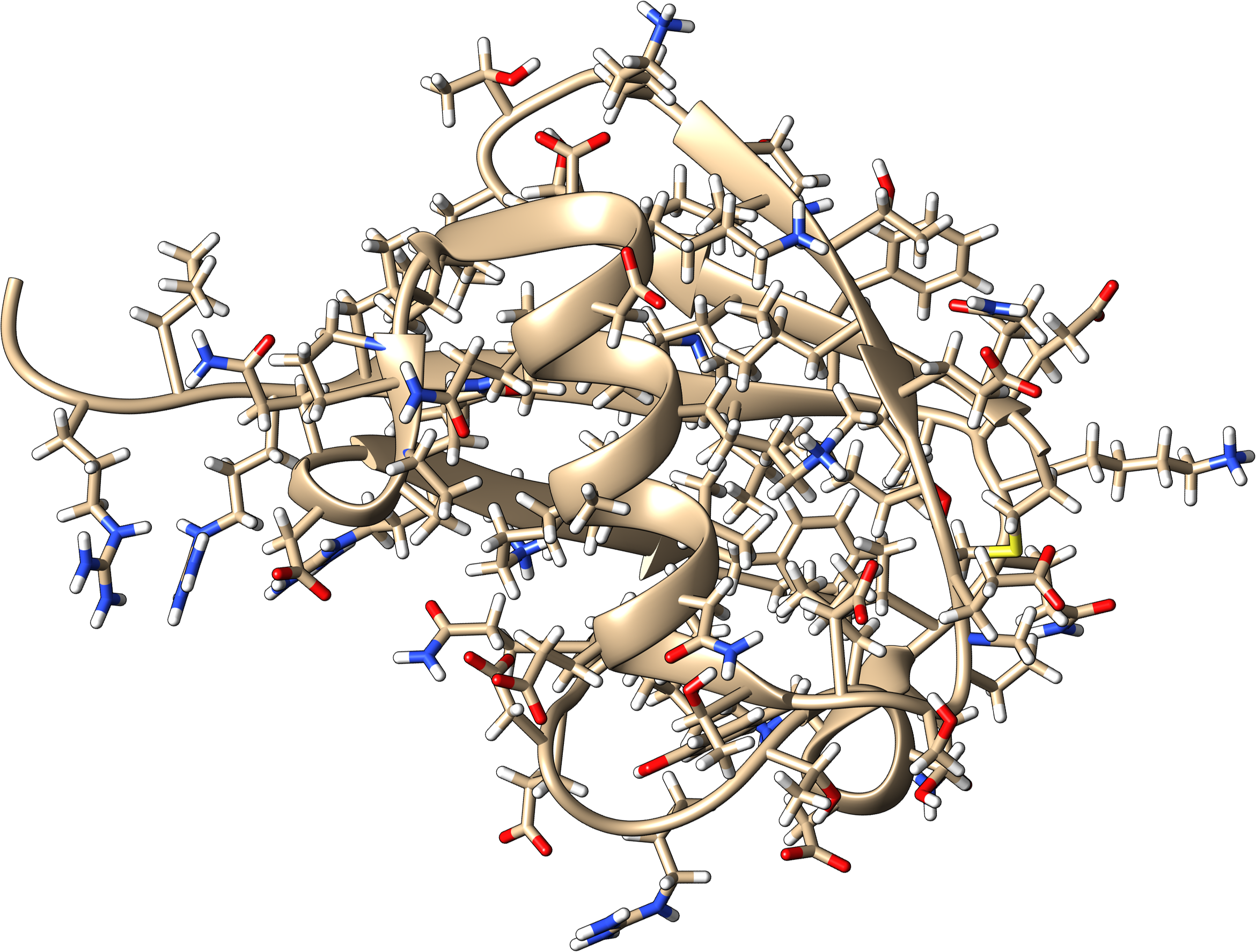}
\caption{Human ubiquitin protein (PDB code 1D3Z): 76 amino acids, 563 magnetic nuclei in the extended backbone (H, N, C, CA, CB, HA, HB)}
\label{fig:ubiq}
\end{figure}
Fully \cC~and \cN--labelled protein human ubiquitin (PDB code 1D3Z, Figure~\ref{fig:ubiq}) containing over a thousand magnetic nuclei in 76 amino acid residues was chosen for testing purposes with two types of spin subsystem selection:
   \emph{backbone} (H, N, C, CA, HA) and
   \emph{extended backbone} (H, N, C, CA, CB, HA, HB).
Both cases involve a weakly branched continuous chain of spin-spin couplings and are encountered in the simulation of a large class of protein backbone NMR experiments that map out the protein bonding network and thereby assist in molecular structure determination:
   HNCO~\cite{kay-nmr3-1990},
   HNCOCA~\cite{bax-nmr3-1991},
   HNCA~\cite{kay-nmr3-1990},
   and HSQC~\cite{nmr-hsqc-1994}.
The isotropic NMR Hamiltonian was assembled using
  chemical shift values from the BMRB database~\cite{biomagresbank-2008}
  and $J$-couplings from the literature data~\cite{case-scalar-2000,lohr-coupling-1999,perez-karplus-2001,vogeli-backbone-2007,wang-dihedral-1996}.
In the cases where an experimental value of a particular $J$-coupling was not available in the literature, it was estimated based on the known values for structurally similar substances~\cite{barfield-allylic-1976,hansen-CH-1981,krivdin-CC-1991} ---
  for most NMR simulation purposes and certainly for the purpose of the demonstration of the performance of the tensor train algorithm the accuracy of such coupling estimates (about $20\%$) is sufficient.
The raw data for the magnetic couplings used in this work is available in the example set supplied with the current public version of the \emph{Spinach} library~\cite{spinach-2011}.

NMR experiments were performed at $25^\circ\mathrm{C}$ on a Varian Inova $600$ MHz ($14.1$ Tesla) spectrometer
   equipped with a Z--gradient triple--resonance cryogenic probe using a $0.5$ mM sample of uniformly \cC~-- and \cN--labelled human ubiquitin in $10\%$ \ce{D_2O}.
\cN~spectra were collected as 2D \cH--\cN~HSQC~\cite{bodenhausen-abundance-1980} spectra incorporating gradient enhanced coherence selection~\cite{kay-single-1992} and water flip-back.
The spectra were recorded with acquisition times of $150 \mathrm{ms}$ ($t_1$, \cN) and $500 \mathrm{ms}$ ($t_2$, \cH).
During the \cN~evolution period, \ce{^1J_{HN}} and \ce{^{1,2}J_{NC}} couplings were
   either allowed to evolve,
   or decoupled by insertion of a rectangular \cN { } or a shaped $200 \mu\mathrm{s}$ \cC~inversion pulse using the central lobe of the sinc function.
During \cH~acquisition \cN~nuclei were either evolved or decoupled using $40$ ppm broadband WURST sequence~\cite{kupce-wurst-1995}.

The liquid state NMR Hamiltonian of \cC,\cN--labelled ubiquitin is:
\begin{equation}\label{eq:H}
 \begin{split}
 \hH(t)  & = \sum_k \omega_k \hS^{(k)}_z
   \\ &        + 2\pi \sum_{l>m} J_{\mathrm{strong}}^{(l,m)} \hvS^{(l)} \cdot \hvS^{(m)}
           + 2\pi \sum_{p>q} J_{\mathrm{weak}}^{(p,q)} \hS^{(p)}_z \hS^{(q)}_z
   \\ &        + \omega_x(t) \sum_r \hS^{(r)}_x
           + \omega_y(t) \sum_s \hS^{(s)}_y
 \end{split}
\end{equation}
where
  canonical NMR spectroscopy notation is used~\cite{ernst-nmr-1987},
  $k$ index runs over all nuclei,
  $l$ and $m$ indices run over pairs of nuclei that belong to the same isotope,
  $p$ and $q$         run over pairs of nuclei that belong to different isotopes,
  $r$ and $s$         run over the nuclei influenced by radiofrequency pulses,
  $\omega_x(t)$ and $\omega_y(t)$ are time profiles of those pulses,
  $\omega_k$ are offset frequencies arising from the chemical shielding of the corresponding nuclei~\cite{ramsey-shield-1950},
  $J_{\mathrm{strong}}^{(l,m)}$ are ``strong'' NMR $J$-couplings~\cite{ramsey-spincoupling-1953},
  $J_{\mathrm{weak}}^{(p,q)}$   are ``weak''   NMR $J$-couplings~\cite{ernst-nmr-1987},
  and spin operators $\hSn_x,\hSn_y,\hSn_z$  are defined by Eq.~\eqref{eq:Sn}.
In the case of extended ubiquitin backbone, the Hamiltonian in Eq.~\eqref{eq:H} contains
  $563$ shielding terms,
  $1840$ coupling terms, and
  $1126$ radiofrequency terms.
All calculations reported below were performed by extending the functionality of \emph{Spinach} library~\cite{spinach-2011}
  to the tensor train formalism and interfacing it to \emph{TT-Toolbox}~\cite{tt-toolbox} where appropriate.

Due to the abundance of complicated multi-pulse NMR experiments with time-dependent Hamiltonians~\cite{ernst-nmr-1987},
  magnetic resonance simulations are generally carried out in the time domain.
They always require long-term evolution trajectories with accurate phases (at least $100$ ms, much longer than the reciprocal Hamiltonian norm) for the density operator $\hrho(t)$  under the Liouville--von Neumann equation:
\begin{equation}\label{eq:LvN}
 \begin{split}
  \frac{d}{dt}\hrho(t) & = -\im \left[ \hH(t),\hrho(t) \right] + \hhR \left( \hrho(t)-\hrho_{\mathrm{eq}} \right), \\
  O(t) & = \left\langle \hO \,\Big|\, \hrho(t) \right\rangle = \trace\left[ \hO^\dagger \hrho(t) \right], \\
  \hrho_{\mathrm{eq}} & = \frac{ \exp \left( - \hH / k_B T \right) }{\trace\exp \left( - \hH / k_B T \right) }.
 \end{split}
\end{equation}
where
  $\hhR$ is the relaxation superoperator ($T_{1,2}$  model with literature values for relaxation times~\cite{cavanagh-nmr-2007} was used in the present work),
  $\hrho_{\mathrm{eq}}$ is the thermal equilibrium state, and
  $\hO$ is the observable operator,
     usually a sum of $\hS_x$  or $\hS_y$ operators on the spins of interest.
In very simple cases where the Hamiltonian is not time-dependent, the general solution to Eq.~\eqref{eq:LvN} can be written as:
\begin{equation}\label{eq:Ot}
 O(t) = \left\langle \hO \,\Big|\, \exp\left[-\im (\hhH+\im\hhR)t \right] \,\Big|\, \hrho_0 \right\rangle,
\end{equation}
where
  $\hhH$ is the Hamiltonian commutation superoperator.

Direct time domain evaluation of this equation in tensor train format, either using explicit operator exponentiation or Krylov type propagation techniques, does not appear to be possible --- in all cases described by Eq.~\eqref{eq:H} the ranks in the tensor train expansion quickly grow beyond the capacity of modern computers.
Increasing the singular value cut-off threshold at the representation compression stage leads to catastrophic loss of accuracy.
Fortunately, there are simple cases (most notably pulse-acquire 1D NMR spectroscopy) where amplitudes at only a few specific frequencies are actually required for the Fourier transform of Eq.~\eqref{eq:Ot}, meaning that the problem can be reformulated in the frequency domain:
\begin{equation}\label{eq:Oo}
 O(\omega) = -\im \left\langle \hO \,\Big|\, \left( \hhH+\im\hhR + \omega\Id \right)^{-1}  \,\Big|\, \hrho_0 \right\rangle.
\end{equation}
That is, to compute the observable at the point $\omega$  in the frequency domain, we need to solve a linear system
$
\left( \hhH+\im\hhR + \omega\Id \right) \left| \hat x \right\rangle = \left| \hrho_0 \right\rangle.
$
%For this operation the efficient DMRG/TT algorithms have recently emerged~\cite{jeckelmann-dmrgsolve-2002,ds-amr1-2013,ds-amr2-2013}.
The problem formulation in Eq.~\eqref{eq:Oo} sacrifices a great deal of generality compared to Eq.~\eqref{eq:LvN} (simulation of arbitrary NMR pulse sequences is no longer possible),
  but it does serve as a stepping stone and enables the demonstration calculation presented below.
%Wider applications of tensor structured methods in magnetic resonance are currently struggling
%  because of large tensor ranks of time-domain solutions .

\section{Tensor train algorithm for the summation and solution of linear systems}  %%%%%%%%%% AMEn
The DMRG algorithm was initially proposed~\cite{white-dmrg-1992,white-dmrg-1993} to find the ground state of a Hermitian matrix $\mA$ by the minimization of the Rayleigh quotient
$
Q(x)= x^* A x/x^*x.
$
The {dynamical} DMRG algorithm~\cite{jeckelmann-dmrgsolve-2002} was then developed to find the solution of a linear system
$
\mA x = b
$
with a Hermitian positive definite matrix $\mA$ by the minimization of the energy function
$
J(x)=x^* A x - 2\Re (x^* b).
$
Apart from the change of the minimization target function, the two algorithms are similar.

In DMRG formalism the solution is sought in the form of a tensor train introduced in Eq.~\eqref{eq:tt},
but the minimization over all cores $x^{(n)}$  simultaneously is a complicated non-linear problem.
To make the procedure feasible, it is replaced by a sequence of optimizations carried over one core at a time:
\begin{equation}\label{eq:local}
 x^{(n)}_\tru = \arg\min_{x^{(n)}} J(\tau(x^{(1)},\ldots, x^{(n)}, \ldots,x^{(N)})).
\end{equation}
The TT format is linear in all cores $x^{(n)}$. This fact may be expressed as $x=X_{\neq n} x^{(n)},$
where the frame matrix $X_{\neq n}$  maps the parameters of the TT core $x^{(n)}$  to the vector $x$.
The linearity allows to rewrite Eq.~\eqref{eq:local} as
$
x_\tru^{(n)} = \arg\min J_n(x^{(n)}) = \mA_n^{-1} b_n,
$
where
 $J_n$ is the energy function for the \emph{local problem} $\mA_n x^{(n)} = b_n$
 with $\mA_n = X_{\neq n}^* \mA X_{\neq n}$
 and $b_n = X_{\neq n}^* b$.
Using the non-uniqueness of the tensor train representation~\eqref{eq:tt}, 
  one can always construct the representation with the unitary frame matrix $X_{\neq n}$, 
  that guarantees the stability of the local problem.
Such a choice is known 
  as \emph{gauge condition} in the MPS literature, 
  and \emph{canonical form} in the DMRG literature.
After the solution $x_\tru^{(n)}$ is computed,
  we substitute $x^{(n)}:=x^{(n)}_\tru$ in the tensor train,
  and continue for $n=1,\ldots,N,$  and then back and forth along the chain.

The convergence of the above described \emph{one-site} DMRG procedure depends on the initial guess and in particular on the initial choice of the TT ranks because they remain the same during the sequence of updates defined by Eq.~\eqref{eq:local}.
This is a severe restriction and additional measures are therefore taken to \emph{adapt} the TT ranks during the computations. 
One way to do that is to replace the optimization over single cores by the optimization over pairs of neighboring cores, and then to adapt the TT rank between them. 
Another possibility is to expand the search space by adding auxiliary directions. 
The first method of the latter type is the \emph{corrected one-site} DMRG algorithm~\cite{white-dmrg1c-2005}, which targets in addition to $x$ a surrogate of the next Krylov vector $Ax.$

For the solution of linear systems, the \emph{alternating minimum energy} (AMEn) algorithm was recently proposed~\cite{ds-amr1-2013,ds-amr2-2013}, which also uses an additional direction to adapt tensor train ranks.
The local optimization step in AMEn is carried over one site only.
To adapt TT ranks and improve convergence, TT blocks are expanded by auxiliary information,
$
x^{(n)} := \begin{bmatrix} x_\tru^{(n)} & r^{(n)} \end{bmatrix}.
$
The \emph{enrichment} $r^{(n)}$ introduces new directions in the subspace spanned by $X_{\neq n+1}$.
A good choice of the enrichment is the component $r^{(n)}$ of the TT representation (exact or approximate)
$
\tilde r = \tau(r^{(1)},\ldots,r^{(N)})
$
of the residual 
$
r = b - Ax.
$
AMEn algorithm is as fast as one-site methods, but as rank adaptive as the two-site DMRG algorithm, 
  and demonstrates comparable or better convergence rates.
For the solution of a linear system $Ax=b$  with a Hermitian positive definite matrix, it has a proven global bound on the geometrical convergence rate.
Unlike the corrected one-site DMRG method~\cite{white-dmrg1c-2005}, 
  the AMEn algorithm is stable to perturbations and free from tuning parameters and heuristics~\cite{ds-dmrgamen-2014}.
The rank adaptation strategy in the enrichment phase of AMEn is determined by a single relative accuracy parameter.

In this work we use the AMEn algorithm for two purposes.
First, we apply it to a system with a trivial matrix $\mA=\Id,$ but a complicated right-hand side $b$, which is a sum of many elementary tensors like the one in Eq.~\eqref{eq:Sn}.
This allows us to compress a Hamiltonian returned by the \emph{Spinach} package from the CP format given by Eq.~\eqref{eq:H} into the TT format Eq.~\eqref{eq:tt}.
The Hamiltonian is stretched into a vector, and
   the target functional $J(x)=\|x-b\|^2$ is a Frobenius-norm distance between a given Hamiltonian $b$ and Hamiltonian $x$ sought in the tensor train format.
The one-site optimization in Eq.~\eqref{eq:local} is effectively the solution of the over-determined linear system $X_{\neq n}x^{(n)}=b$ using the least squares method.
For the unitary frame matrix we have $x^{(n)}_\tru=X_{\neq n}^* b,$
  and therefore the local optimization step is obtained by contracting the frame matrix with the given Hamiltonian $b.$
The enrichment step uses a low-rank approximation of the error $r=b-x,$
  which is obtained by one-site DMRG optimization.

After the Hamiltonian is compressed into the tensor train format, we compute 1D NMR spectra by solving the linear system in Eq.~\eqref{eq:Oo}.
Since the matrix 
$
\hhA = \hhH+\im\hhR+\omega\Id
$ 
is not expected to be Hermitian positive definite, we consider instead an equivalent symmetrized problem
$
(\hhA^* \hhA) \hrho(\omega) = \hhA^* \hrho_0.
$
For demonstration purposes, we chose a simple non-selective damping relaxation model
$
\hhR=-\mu\Id,
$
and the same $\hS_{+} = \hS_x + \im\hS_y$ operator for the initial and the detection state,
where $\hS_{\{x,y\}}=\sum_n\hSn_{\{x,y\}}$ are the total spin operators of all \cN~nuclei in the system.
This avoids explicit radiofrequency pulses and makes the Hamiltonian in Eq.~\eqref{eq:H} time-independent and real-valued, 
$
\hH(t)=\hH = \hH^* = \hH^T,
$
those properties are also inherited by the commutation superoperator $\hhH.$
Since the detection state $\hO$ is also real-valued, 
 the NMR spectrum $O(\omega)$ in Eq.~\eqref{eq:Oo} can be computed from $\Im\hrho(\omega),$ 
 that we obtain as follows:
\begin{equation}\label{eq:rho}
 \Im \hrho(\omega) = \mu \left( \hhH^* \hhH + 2\omega \hhH + (\omega^2+\mu^2)\Id \right)^{-1} \hrho_0.
\end{equation}
This equation is solved by the AMEn algorithm at each point $\omega$ in the user-specified frequency interval.

\section{Results} %%%%%%%%%%%% RESULTS
\begin{figure}[t]
 \includegraphics[width=\linewidth]{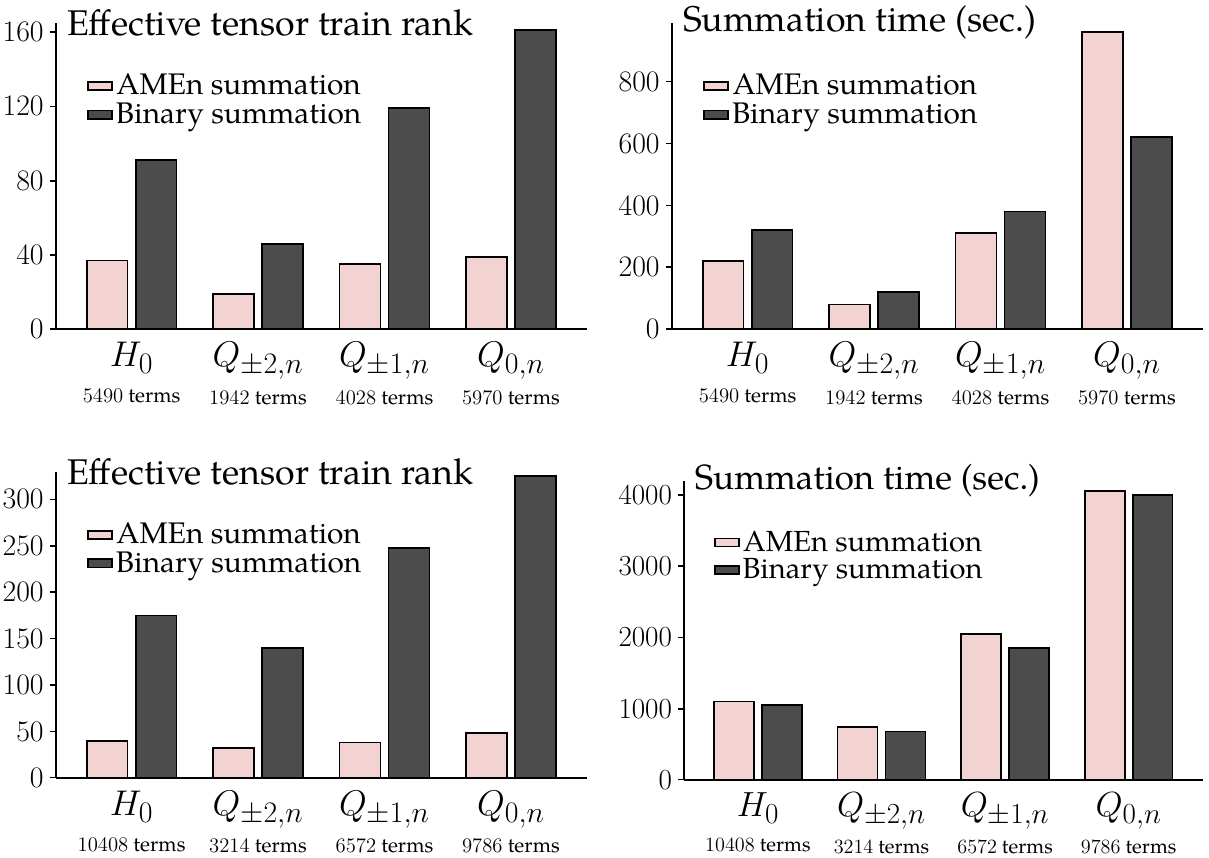}
\caption{Performance comparison for binary and AMEn tensor train summation with relative accuracy parameter $\eps=10^{-12}$ during the construction of the NMR spin Hamiltonian.
\textbf{Top} ---  human ubiquitin backbone (H, N, C, CA, HA),
\textbf{bottom} --- human ubiquitin extended backbone (H, N, C, CA, HA, CB, HB).
Here $H_0$  refers to the isotropic part of the Hamiltonian
and $Q_{k,n}$ to the irreducible spherical components of the anisotropic part.}
\label{fig:time}
\end{figure}

As discussed above, a major problem in the application of tensor train methods to magnetic resonance simulation of large systems is the calculation of lengthy sums involved in the construction of spin Hamiltonians and density matrices, and their compression into the TT format.
Fig.~\ref{fig:time} illustrates the performance of our proposed solution to this problem in the case of
  minimal (H, N, C, CA, HA) and
  extended (H, N, C, CA, HA, CB, HB) ubiquitin backbone spin systems.
Storage requirements for the TT format in Eq.~\eqref{eq:tt} depend on all TT ranks (bond dimensions) $k_1,\ldots,k_{N-1},$
and are characterized by the \emph{effective} TT rank $k,$ defined by
$
N k^2 = \sum_{n=1}^N k_{n-1} k_n.
$
It is clear from the left panels of Fig.~\ref{fig:time} that the primary showstopper
   --- rapid growth in the tensor train rank ---
   has been removed by the AMEn method:
  the effective ranks stay below $50$ for the extended backbone and below $40$ for the minimal backbone,
  well within the capability of modern desktop workstations.
Since $k^2$ is smaller than the number of terms in the CP representation, the TT format with $Nk^2$ operators provides more compact storage than the CP format.

The alternative to AMEn is \emph{binary} summation, which adds up Hamiltonian terms pairwise and recompresses the representation after each addition. As demonstrated in Fig.~\ref{fig:time}, binary summation drives tensor train ranks up to several hundred and thereby makes the solution of the linear system in Eq.~\eqref{eq:rho} exceedingly difficult.
It is clear from the right panels of Fig.~\ref{fig:time} that the CPU time requirements of AMEn summation compared to binary summation are essentially the same, making AMEn procedure clearly superior for all practical purposes. The resulting representation of the ubiquitin backbone spin Hamiltonian matrix is, up to the rounding error of the complex double precision arithmetic, exact.
In magnetic resonance spectroscopy this is an unprecedented development --- ubiquitin NMR simulation is currently just about feasible~\cite{south-ubiq-2014}, with significant approximations and colossal computational resources.
Tensor train representation is therefore a large step forward,
  even though Eq.~\eqref{eq:Oo} is not in general applicable to arbitrary NMR experiments.
\begin{figure}[p]
 \includegraphics[width=\linewidth]{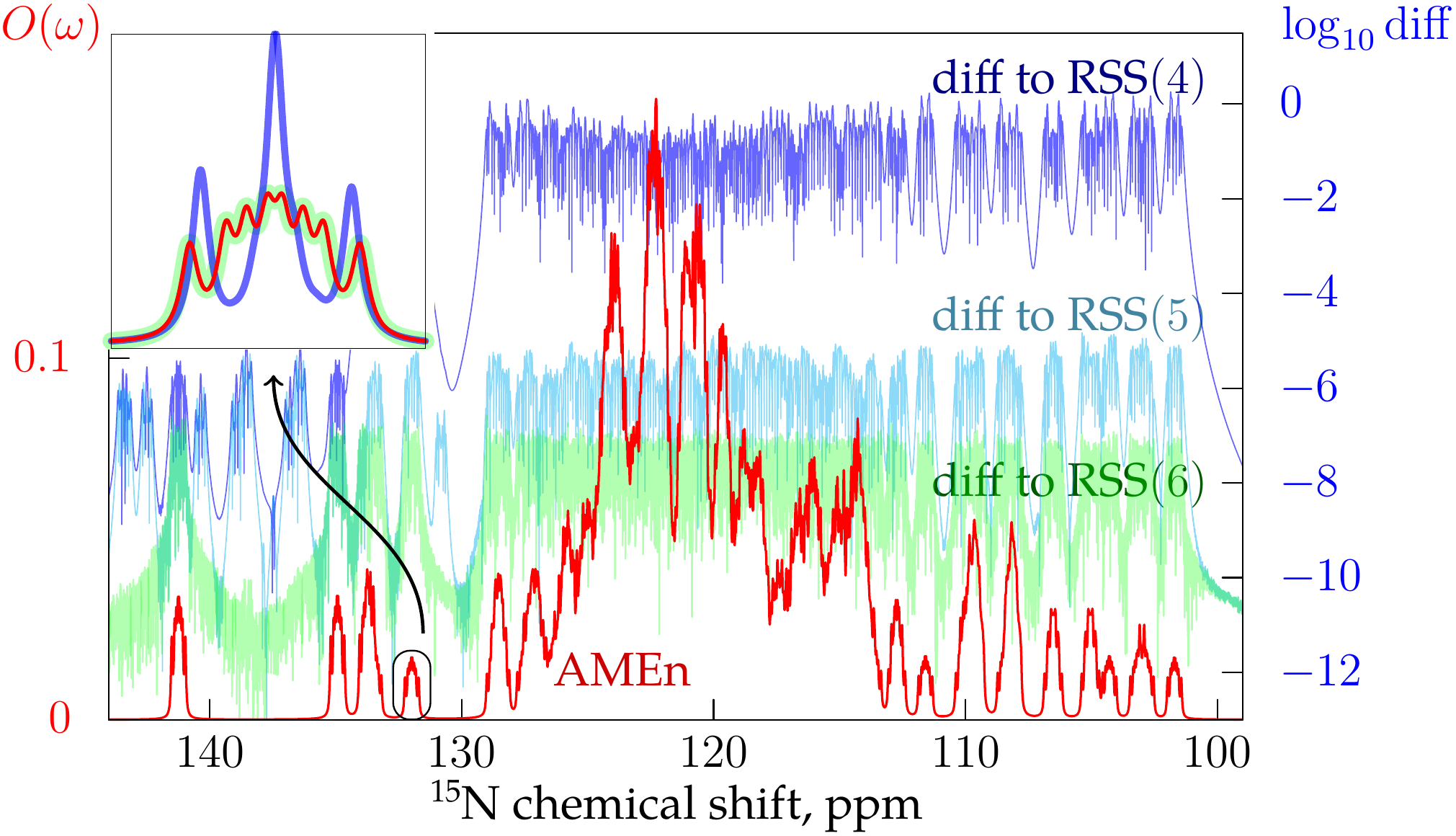}
 \vskip 3mm
 \includegraphics[width=\linewidth]{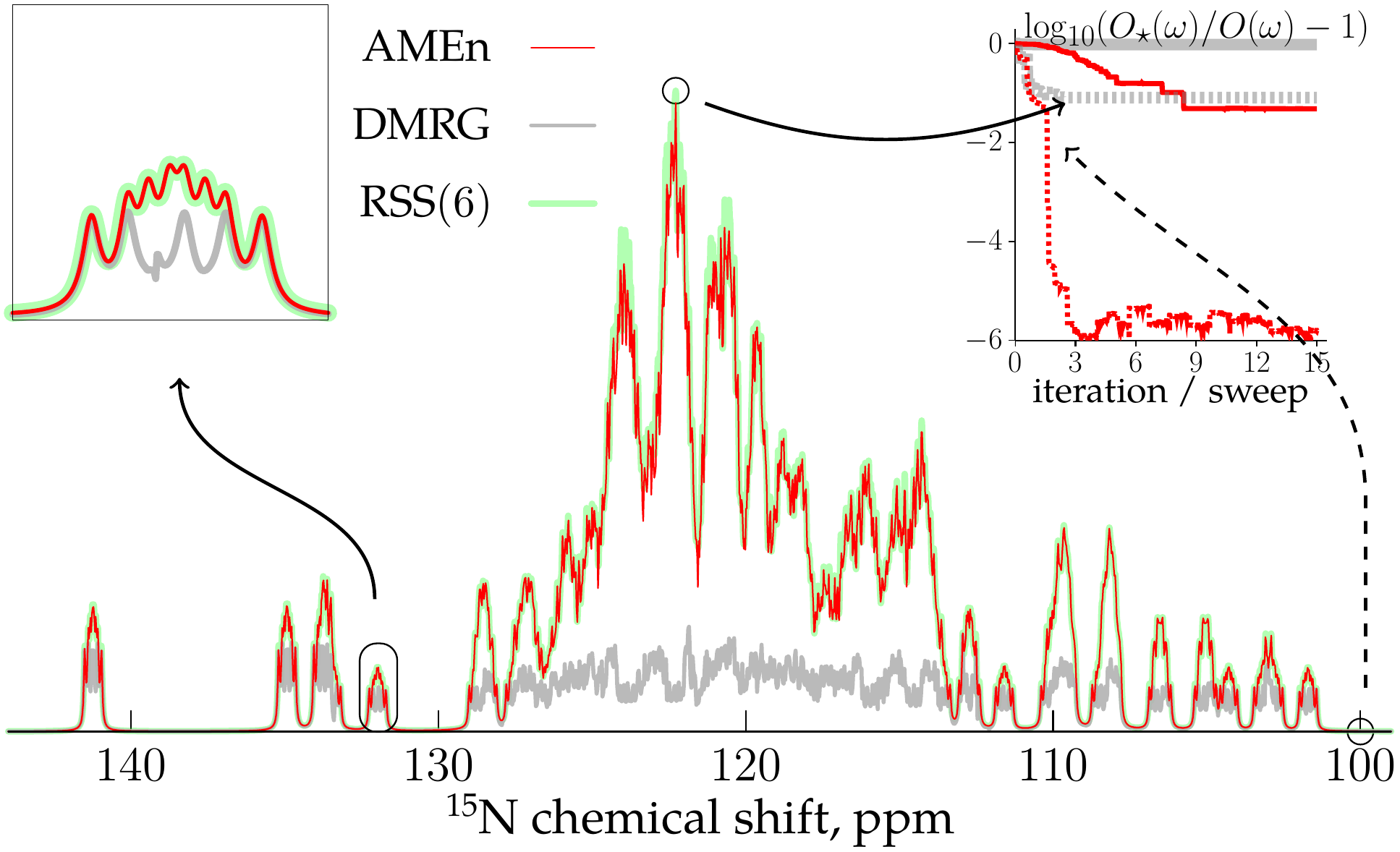}
\caption{Amino group region of the pulse-acquire \cN~NMR spectrum of human ubiquitin.
\textbf{Top} --- 
  spectrum computed by AMEn algorithm~\cite{ds-amr1-2013,ds-amr2-2013} with accuracy parameter $\eps=10^{-6}$ 
  is compared to the results obtained by the restricted state space (RSS) approximation~\cite{spinach-2011} 
  with basis containing local spin correlations of orders up to $5$ and $6$. 
\textbf{Bottom} --- accurate RSS computation is used as a reference $O_\star(\omega)$ and compared to the spectra $O(\omega)$ computed by AMEn and DMRG~\cite{jeckelmann-dmrgsolve-2002,DoOs-dmrg-solve-2011}, both using the accuracy parameter $\eps=10^{-3}.$
Right subgraph: convergence of AMEn and DMRG methods at two points of the frequency domain (dashed lines: an off-peak point at $100$ ppm, solid lines: a peak at $122$ ppm).
}
\label{fig:1d}
\end{figure}
\begin{figure}[p]
 \hbox to \textwidth{\hfill
 \includegraphics[height=.38\textheight]{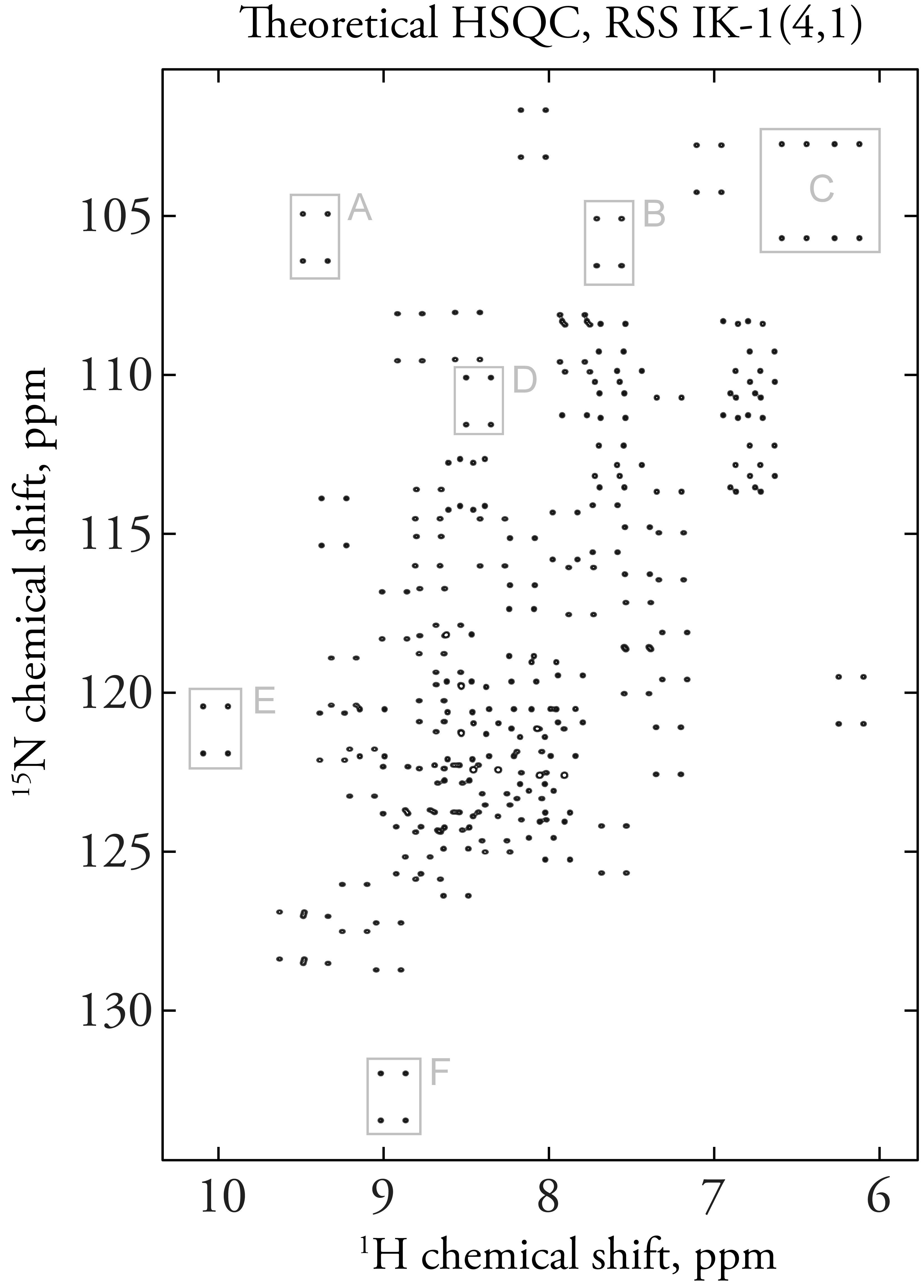} \hfill
 \includegraphics[height=.38\textheight]{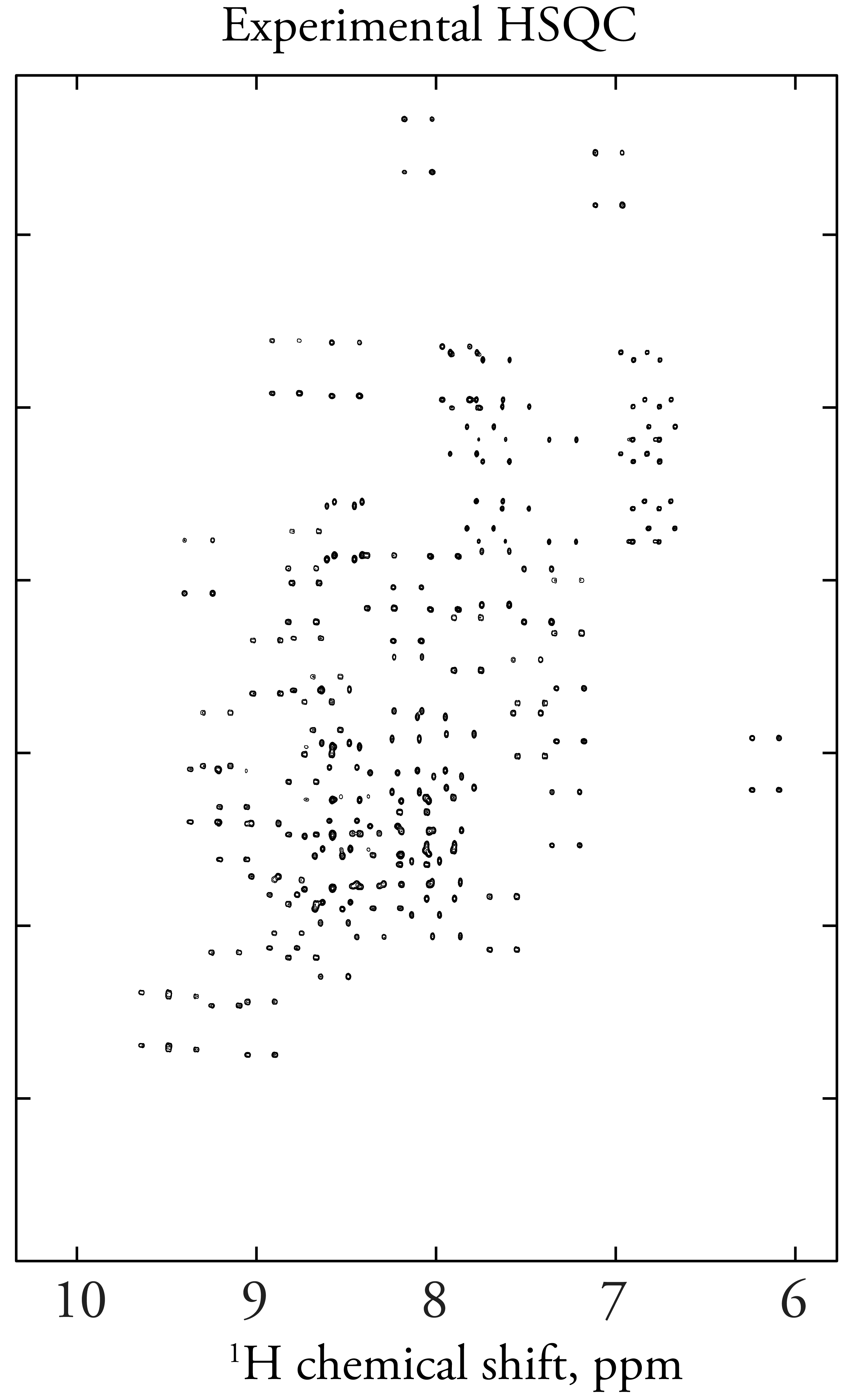} \hfill
 }
 \vskip 7mm
 \hbox to \textwidth{\hfill
 \includegraphics[height=.38\textheight]{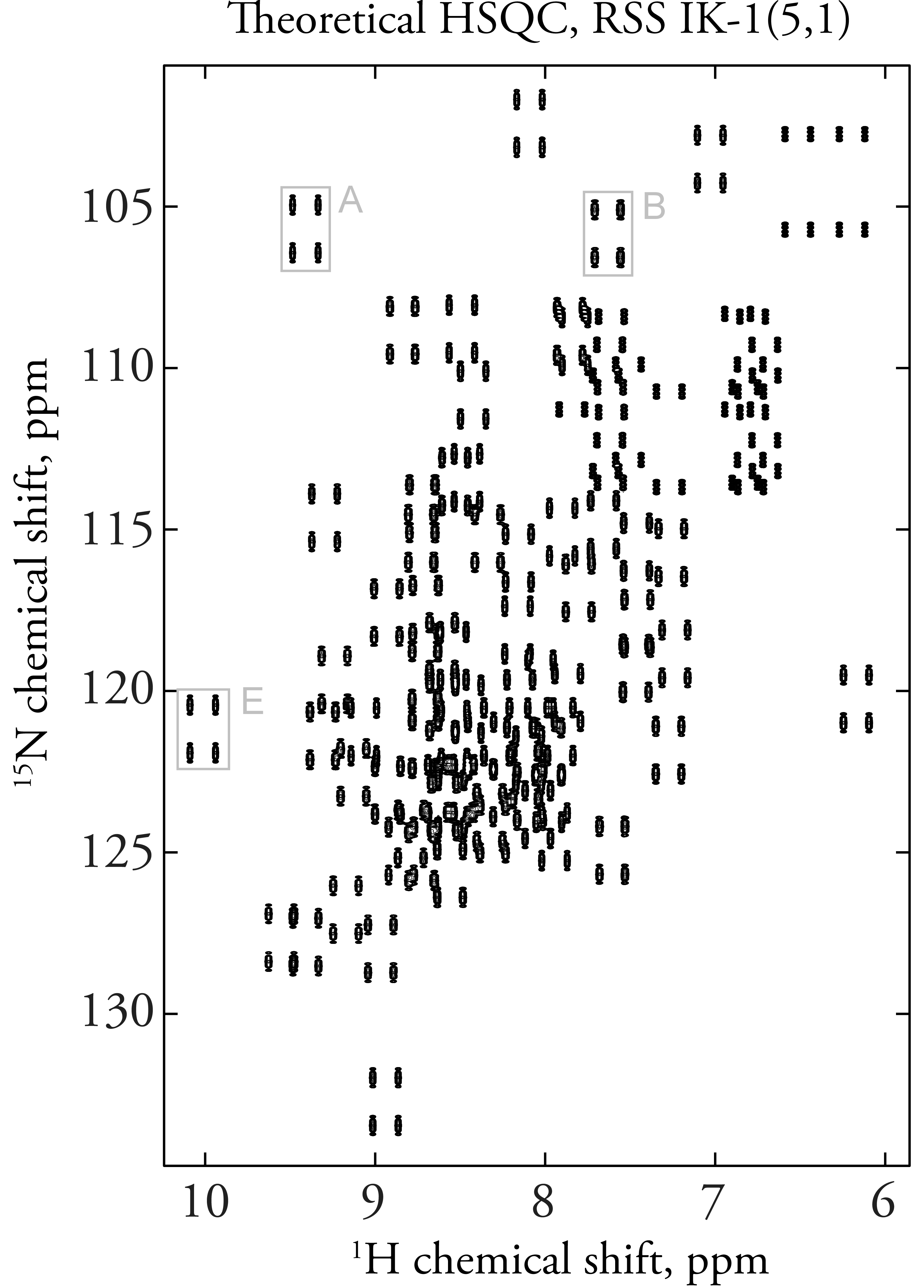} \hfill
 \includegraphics[height=.38\textheight]{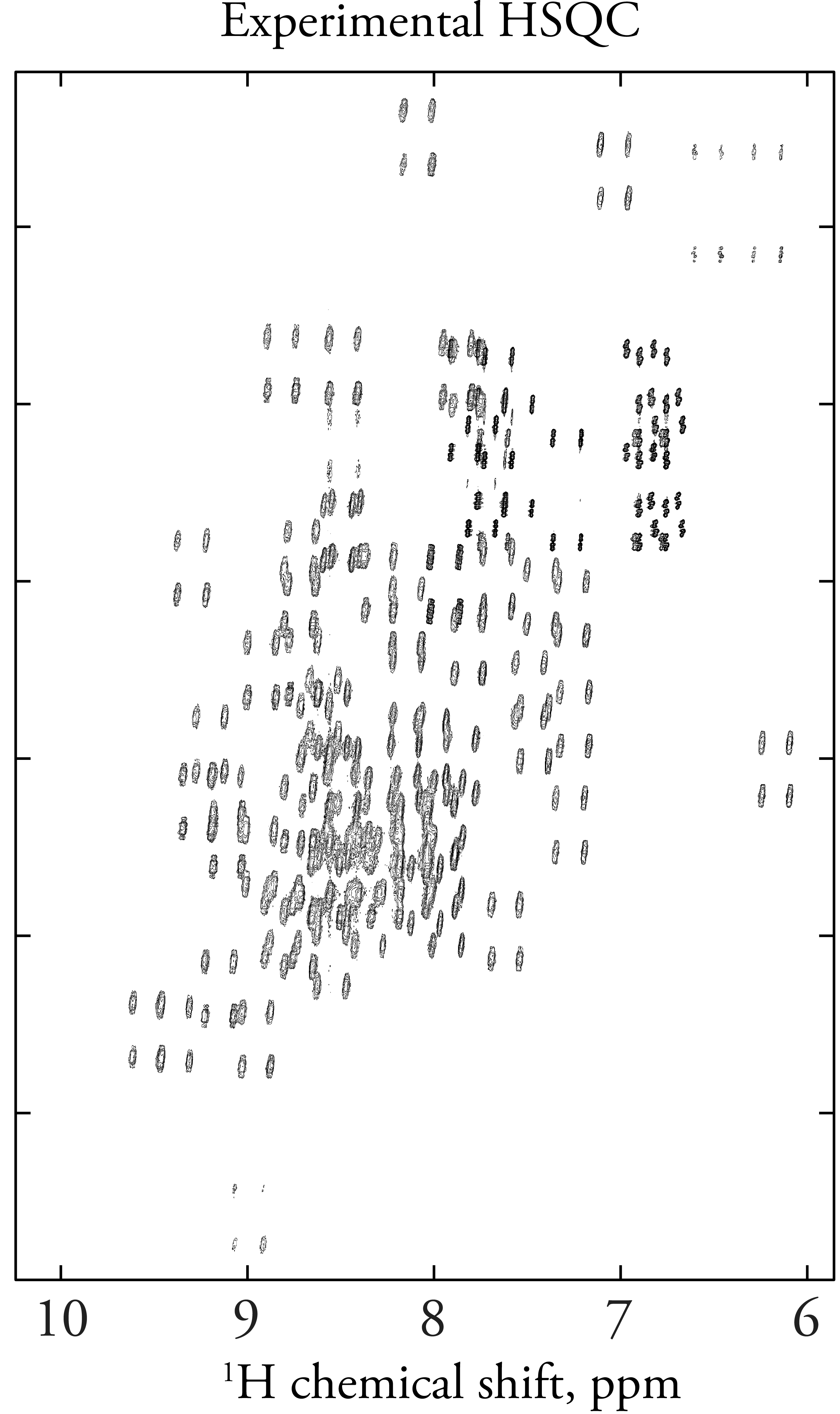} \hfill
 }
\caption{Theoretical (\textbf{left}) and experimental (\textbf{right}) \cH-\cN~HSQC spectra of \cN,\cC--labelled human ubiquitin.
\textbf{Top} --- proton decoupling switched off in the indirect dimension and nitrogen decoupling switched off in the direct dimension to demonstrate accurate quantum mechanical treatment of spin-spin coupling by the simulation.
\textbf{Bottom} --- additionally, carbon decoupling is switched off in the indirect dimension.
Signal groups marked A-F in the theoretical spectrum are not visible in the experimental data due to partial deuteration and slow conformational exchange of the corresponding amino acid residues.}
\label{fig:2d}
\end{figure}

After the Hamiltonian is compressed, we compute \cN~pulse-acquire NMR spectra using Eq.~\eqref{eq:rho} with the AMEn algorithm and compare it to the simulation produced by the restricted state space (RSS) approximation~\cite{spinach-2011}, which is currently the only other method that is capable of handling NMR systems of this size.
As demonstrated in Fig.~\ref{fig:1d} (top), 
When the basis set used by RSS is increased, its result converges to the one produced by AMEn,
 and the relative deviation between two methods falls below $10^{-6}$ across the frequency interval.

It is instructive to compare the results of AMEn simulations with those produced by the dynamical DMRG~\cite{jeckelmann-dmrgsolve-2002} technique. As shown in Fig.~\ref{fig:1d} (bottom), the NMR spectrum computed by AMEn matches the reference spectrum returned by RSS with only minor deviations, while the accuracy of the result computed by the dynamical DMRG algorithm at the same relative accuracy parameter is unacceptable. DMRG does of course produce the right answer if a much tighter accuracy parameter is specified, but the simulation time goes up by several orders of magnitude. AMEn does therefore appear to have a better accuracy-to-effort ratio. This is also confirmed by the convergence graph of AMEn and DMRG, given in the same figure, where the relative deviation between the computed and the reference values $O(\omega)$ is shown during the iterations (sweeps) for both DMRG and AMEn. Note also that the inexact values of the spectrum, computed by AMEn and DMRG are always below the reference values; this was first noted by Jeckelmann~\cite{jeckelmann-dmrgsolve-2002}. The comparison in Fig.~\ref{fig:1d} is made using $\eps=10^{-3}$ to visually emphasize the observed difference between the two methods; the same conclusion also holds for more accurate calculations using $\eps=10^{-8}.$

Due to the intrinsically low sensitivity of liquid state \cN~protein NMR spectroscopy, it is not possible to record the experimental equivalent of Fig.~\ref{fig:1d} directly with a sufficient signal-to-noise ratio; we have therefore taken a somewhat longer route to the experimental validation of the tensor train simulation --- Fig.~\ref{fig:2d} shows experimental proton-detected \cN--\cH~HSQC spectra of ubiquitin, compared to the simulations obtained at the basis set limit of the RSS formalism~\cite{karabanov-acc-2011}. Perfect agreement is apparent in both cases. This provides an experimental evidence to the accuracy of the restricted state space method. The tensor train results in Fig.~\ref{fig:1d} can now be justified by comparison to the RSS results --- it is clear that the TT formalism performs as intended.

\section{Discussion} %%%%%%%%%%%% DISCUSSION

The successful 1D NMR simulation notwithstanding, very significant obstacles remain on the path to practical applications of the tensor train formalism to NMR spectroscopy. The following issues should be addressed in future work to fully uncover the potential of the DMRG/MPS/TT formalism for spin dynamics simulations:

(a) The requirement for the spin system to be a chain or a tree should be lifted. Biological magnetic resonance spin systems are irregular polycyclic interaction networks with multiple interlocking loops in the coupling graph, particularly in solid state NMR, where inter-nuclear dipolar couplings form very dense meshes. A generalization of tensor train algorithms to general contraction networks that fully mimic the molecular structure is therefore required.

(b) Rank explosion problem for time-domain simulations should be solved. It is clear from the success of the restricted state space approximation~\cite{kuprov-2007,spinach-2011} that the order of spin correlation in many evolving magnetic resonance spin systems either is or may safely be assumed to be quite low. This suggests data sparsity and separability, and indicates that some kind of low-rank decomposition is possible. One likely direction is through the enforcement of symmetries and conservation laws within the tensor train format itself during time evolution.

(c) Our experience indicates that tensor train objects are very far from being drop-in replacements for their matrix counterparts in standard simulation algorithms and software --- it does actually appear that nearly everything in the very considerable body of magnetic resonance simulation methods needs to be adapted to the realities of DMRG. Current implementation of tensor product methods still requires a number of tuning parameters (approximation accuracies, TT ranks of the enrichment, etc.). Broad adoption of tensor network algorithms would require basic linear algebra operations to be handled transparently and seamlessly by the existing simulation software packages, in the same way as sparse matrices currently are.

(d) Transparent and clear tensor train approximation accuracy criteria, rank control and \emph{a priori} error bounds should be developed in order to estimate the influence of the representation compression errors on the accuracy of the final result. This problem is particularly acute for the state vector phase in time domain simulations: magnetic resonance experiments rely critically on the phase being correctly predicted.

All of that having been said, we are very optimistic about the future of low-rank tensor product DMRG/MPS/TT methods, having also found them useful in Fokker--Planck type formalisms related to NMR and EPR spectroscopy~\cite{south-nmr-2013}. Their primary strength is the lack of heuristic assumptions and the controllable nature of the representation accuracy. An experimental implementation of tensor train magnetic resonance simulation paths, via an interface to the \emph{TT-Toolbox}~\cite{tt-toolbox}, is available in version 1.3.1980 of our \emph{Spinach} library~\cite{spinach-2011}.

\section{Conclusions and future work}  %%% CONCLUSIONS
Even with their well-documented limitations (the requirement for the spin system to be close to a chain, difficulty with long-range time-domain simulations,
code implementation challenges, etc.), the ability of tensor network formalisms to simulate simple liquid state NMR spectra of large spin systems essentially without approximations is impressive. They cannot yet match the highly optimized dedicated methods developed by the magnetic resonance community~\cite{south-ubiq-2014}, but if some of the limitations are lifted by the subsequent research, DMRG methods would have the potential to become a very useful formalism in NMR research.

Having solved in this paper the last purely technical problem on the way to the broad adoption of tensor train formalism in magnetic resonance spectroscopy, we are quite optimistic about its potential. In particular, the following avenues appear promising:
\begin{enumerate}
\item Generalizing AMEn method to arbitrary tensor networks, e.g. tree tensor networks~\cite{vidal-ht-2006,verstraete-ht-2010,verstraete-ht-2013,hk-ht-2009}, that closely match the coupling topology of the spin system.
\item Development of reliable tensor train methods for solving linear systems of algebraic equations with indefinite matrices, and time evolution problems.
\item Development of tensor product methods that reduce memory requirements and accelerate convergence by enforcing conservation laws~\cite{Ramasesha-symm-dmrg-1996,mcculloch-nonabel-dmrg-2002,schollwock-2005} and matrix symmetries~\cite{huckle-symmetries-2013}.
\end{enumerate}
Elsewhere in magnetic resonance, benefits to electron spin resonance spectroscopy, with its star-shaped spin interactions graphs, are likely to be harder to achieve, but may still be obtained by exploiting the direct product structure of combined spin and spatial dynamics appearing in Fokker--Planck type problems~\cite{south-nmr-2013}.

\section*{Acknowledgements}
We are grateful to Garnet K.-L. Chan for patiently explaining the DMRG formalism to IK during his visit to Cornell University and to Zenawi T. Welderufael for finding some of the less obvious ubiquitin $J$-couplings in the literature. We acknowledge the Iridis High Performance Computing Facility, and the associated support services at the University of Southampton. JMW would like to acknowledge the Wellcome Trust for support of the Southampton NMR centre and we would like to thank the Geoff Kelly at NIMR (Mill Hill) for lending us the ubiquitin sample. The project is supported by EPSRC (EP/H003789/2, EP/J013080/1).

%\bibliographystyle{siamdoi}
%\bibliography{dmrg,nmr,our,tensor}   %% BIBLIOGRAPHY

\end{document}